# The origin of Sr segregation at $La_{1-x}Sr_xMnO_3$ surfaces

Walter A. Harrison
Applied Physics Department
Stanford University
Stanford, CA 94305

Abstract

A uniform distribution of La and Sr in lanthanum-strontium manganites would lead to charged crystal planes, a charged surface, and arbitrarily large surface energy for a bulk crystal. This divergent energy can be eliminated by depleting the La concentration near the surface. Assuming an exponential form for segregation suggested by experiment, the total electrostatic energy is calculated, depending only upon the decay length and on an effective charge $Z^*$ associated with the La ion. It is found to be lower in energy than neutralization of the surface by changing Mn charge states, previously expected, and lower than simply readjusting the La concentration in the surface plane. The actual decay length obtained by minimizing this electrostatic energy is shorter than that observed. The extension of this mechanism to segregation near the surface in other systems is discussed.

Dulli, Dowben, Liou, and Plummer [1] have found that strontium segregates near the surfaces of $La_{1-x}Sr_xMnO_3$ (LSM). Their data is replotted as the La concentration, $1-x$, as a function of distance from the surface in Fig. 1. The curve is very much the same for AO surfaces and for $BO_2$ surfaces. It is an important consideration for the use of this material as a cathode in solid-oxide fuel cells since $SrMnO_3$ is insulating and this segregation can make the surface areas of the cathode, where oxygen may be incorporated, insulating. The mechanism for this segregation has not been established. Dulli, et al.[1] also found evidence for structure changes at the surface, but here we shall think in terms of the starting perovskite structure.

We point out here that $SrMnO_3$ has formally neutral (100) planes, alternate AO (or SrO, with +2, –2 formal valences) and $BO_2$ (or $MnO_2$, with +4. –2, –2 formal valences). With a fraction $1-x$ of the Sr ions replaced by La, the AO planes take a charge $+(1-x)$ per area $a^2 = 4d^2$, with $d$=1.96 Å, the Mn-O spacing in the perovskite crystal. In all of the remaining discussion we shall think in terms of charge per area $4d^2$, or per A site, and not always repeat the "per $4d^2$". Then for charge neutrality a fraction $1-x$ of the $Mn^{4+}$ ions in the bulk become $Mn^{3+}$ ions and the $BO_2$ planes take a formal charge $-(1-x)$. It has long been known[2-5] (and will be illustrated in Fig. 2) that terminating such a set of charged planes leaves an effective surface charge of $\pm(1-x)/2$ ( in units of $e$, again per $4d^2$), necessarily either a field in the crystal or outside of the crystal, and arbitrarily large field energy for bulk crystals. If this is a $BO_2$ surface the field directions are opposite to those from an AO surface, but still arbitrarily large.

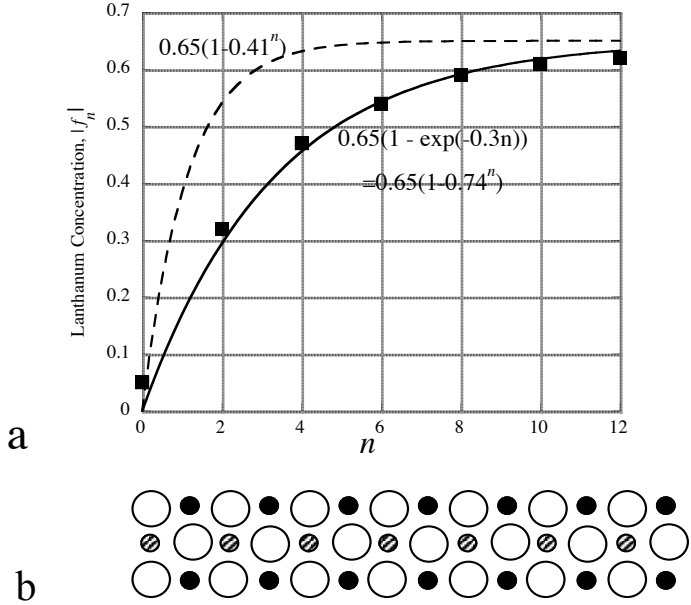

Fig. 1. The squares in Part a are the fraction $f_n$ of A sites occupied by La in (100) planes of $La_{0.65}Sr_{0.35}MnO_3$, as measured by Dulli, et al.[1], for planes numbered $n$ (starting with $n = 0$ for the surface AO plane). The solid line is a fit to the data. The dashed line is the exponential which gave lowest electrostatic energy, with the full Madelung terms of Eq. (B4) included. Part b represents the lattice; striped circles are A atoms, black circles are Mn and open circles are oxygen.

Locally shifting electronic charge between neighboring ions can reduce this surface charge, as could dielectric screening of charge, but in an insulator it cannot eliminate it. It *does* have the effect of reducing the charge of the plane from $\pm(1-x)$ to some $\pm Z^*(1-x)$, as if it was a reduction of all charges by a dielectric constant, $\varepsilon = 1/Z^*$. We may think of $Z^*$ as the effective charge of an La ion, though we see in Appendix B that it is really a $Z^*(La)-Z^*(Sr)$, which we estimate as $Z^*=0.63$. The surface always finds a way for this to be neutralized. When it occurs at an interface, such as between $SrTiO_3$ and $LaAlO_3$, as in Refs. 3 and 4, it is accomplished by introducing electrons in the $SrTiO_3$ conduction bands. We have assumed[6] that for LSM it occurred by modifying the valence charges of Mn ions at the surface, the counterpart of doping but now for a system which is better described[7] in terms of localized cluster states rather than bands. LSM does in fact respond to doping by changing these charge states. An alternative mechanism is suggested by this Sr segregation; the concentration of La could be gradually reduced until it was zero near the surface. Then terminating the crystal with the AO or a $BO_2$ surface would seem to be equivalent and we might expect no field. We explore that mechanism here and find that it can in fact eliminate the surface charge and that it leads to a distribution of La ions qualitatively consistent with that shown in Fig. 1. Because it is accomplished by removal of the charged La ions, we prefer to think of it as a surface depletion of La ions, rather than the equivalent Sr segregation at the surface.

We consider a perovskite crystal with (100) planes as shown in Fig. 1b. As in the Dulli, et al.[1] plot, the planes are numbered $n = 0$ for a surface AO plane, then odd numbered $BO_2$ planes and even numbered AO planes. This represents an AO surface, but simply removing the $n=0$ plane of ions would make it a $BO_2$ surface. We continue with the AO surface and expect that the argument would be changed little for a $BO_2$ surface. The crystal continues to some large $n = N_0$ (with an even number of planes of ions so the total charge will vanish) which we may think of as the center of the crystal. We let the fraction of sites $f_n$ in the even planes be La, and will corresponding let a fraction $f_n$ of the Mn ions in the odd $n$ planes be $Mn^{3+}$ to maintain approximate local charge neutrality. We shall retain exactly the same total number of La ions as we modify the $f_n$ so we have the same total number of $Mn^{4+}$ and $Mn^{3+}$ cluster states (localized states based upon an Mn ion and its surrounding six oxygen neighbors), and the same number of Sr and La cluster states (each with twelve neighboring oxygens), which we treated in detail in Refs. 6 and 7, so the bonding energy does not change with the rearrangement of ions. However, modifying the $f_n$ *does* change the Coulomb interaction energies and those electrostatic terms are what we treat here. For that reason the only materials parameter which enters our results is $Z^*$, in addition to the composition $x$ and the interatomic distance $d$.

In bulk LSM, with a charge $\pm Z^*e(1-x)$ per $4d^2$ on alternate planes, we may use a one-dimensional Poisson's Equation to find alternate fields between planes of $\pm 4\pi Z^*e(1-x)/[2(4d^2)]$, and alternate potential energies of La ions $\pm \pi Z^{*2}e^2(1-x)/4d$, to be multiplied by the fraction of sites in the plane, $f_n = 1-x$, to be twice (because each interaction is counted twice) the electrostatic potential energy per site in the plane. Near the surface we assume the fraction of A sites which are La suggested by the experiments of Dulli, et al.[1], $f_n = (1-x)(1-e^{-\mu dn})$ for even $n$, shown as the solid line in Fig. 1a, rising exponentially from a surface value of zero to the bulk $f_n = 1-x$. With the fraction of $Mn^{3+}$ sites on the odd-$n$ planes following the same formula but of opposite sign, the charge on the $n$th plane is $eZ^*$ times

$$f_n = (-1)^n(1-x)(1-e^{-\alpha dn}) = (-1)^n(1-x)(1-r^n). \tag{1}$$

Writing the exponential decay, $e^{-\alpha dn}$, as $r^n$ simplifies the subsequent formulae. Note this places *no* La ions on the $n = 0$ plane so that it is neutral and has no contribution to the electrostatic energy. Thus we could think of this as an $MnO_2$-terminated surface, beginning at $n = 1$. Indeed with $r = 0$, it would be a uniform crystal, with an effective surface charge of $-(1-x)/2$, negative because with the $n=0$ plane neutral the first charged plane is $BO_2$.

We associate an electrostatic potential $\phi_n$ with each plane, an average over the plane based only on the deviations from neutral planes, so that each plane can be treated as a uniform sheet of charge as for the bulk above. A calculation of these potentials from the assumed form of $f_n$ was made in Appendix A, leading to

$$e\phi_n = (-1)^n (\pi Z^*e^2/d)(1-x)[1/4 - r^{n+1}/(1+r)^2]. \tag{2}$$

This is plotted in Fig. 2 for $r = 0.74$, from Fig. 1. It very nearly goes to zero at $n = 0$ for this choice of $r$, but it would not for much smaller $r$.

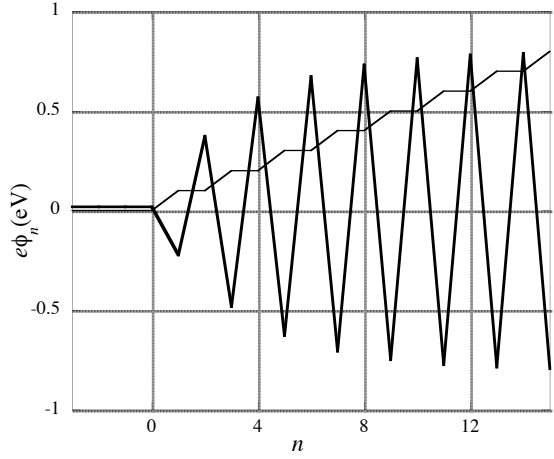

Fig. 2. The potential as a function of plane number, from Eq. (2) with $r=0.74$, as the heavy line. The light line is the corresponding plot often used[2-5] to illustrate a charged surface, with the steady rise representing an internal electric field.

We then evaluated the total electrostatic energy, $\frac{1}{2} \Sigma_n eZ^* f_n \phi_n$, of the ions in these planes as a function of $r$. As $r$ increases from zero, more La ions are eliminated at the surface and these, as well as the corresponding $Mn^{3+}$ ions, were inserted in bulk sites, adding to this electrostatic energy. It was noted in the course of the calculation that this distribution, Eq. (1), eliminates only part of the initial surface charge of $-(1-x)/2$, but we could place $\frac{1}{2}(1-x)(1-r)/(1+r)$ La ions on the $n=0$ plane which Eq. (1) left with no La, and that reduced the surface charge to zero. It is interesting that for the experimental $r = 0.74$ this comes out to be an occupation of $f_0 = 0.049$, as close as one can tell equal to the $n=0$ experimental point in Fig. 1. The number required at $n = 0$ decreases as $r$ increases. [We could instead have promoted that number of Mn ions from $Mn^{3+}$ to $Mn^{4+}$ states but we see at Eq. (4) that it would require a greater energy. Such a promotion would have been consistent with our earlier choice[6] of $\frac{1}{2}(1-x)$ promotions and no La rearrangements (corresponding to taking $r = 0$ in terms of Eq. (1).] The resulting total electrostatic surface energy was found in Appendix A as

$$E_{tot.} = (\pi Z^{*2} e^2/2d)(1-x)^2 \{ -r^2/[(1-r)(1+r)^3] + 1/[4(1-r)] - r(1-r)/(1+r)^3 \}. \qquad (3)$$

The first term is the energy gained by removing La ions from the surface region. The second term is the electrostatic energy cost of transferring these La ions to the bulk, including the $1-x$ ions removed from the surface plane at the outset by setting $f_n=0$. The third is for transfer back of La ions from the bulk to the $n=0$ plane to exactly neutralize the surface.

We note in passing that this last term from transferring $\frac{1}{2}(1-r)(1-x)/(1+r)$ La ions to the surface could be interpreted as an *electrostatic* promotion energy for neutralizing the surface by reducing the number of La ions in the surface plane, given by

$$E_{pro}^{elec.} = (\pi Z^{*2}e^2(1-x)/d)r/(1+r)^2 = 1.46 \text{ eV}. \tag{4}$$

This is smaller than the 2.58 eV estimated in Refs. 6 and 7 for promoting an $Mn^{4+}$ ion to $Mn^{3+}$, supporting the proposition that neutralization comes from La rearrangement rather than from Mn ion promotion.

The numerical estimate was based on the $r = 0.74$ from Fig. 1 and the $Z^* = 0.63$ from tight-binding theory of the electronic structure carried out in Appendix B. These parameters give a leading factor in Eq. (3) of 1.94 eV. $E_{tot}$ is found to be a minimum at $r=0.19$, independent of the value of the leading factor, corresponding to very short penetration of the La depletion region, compared to the $r=0.74$ in Fig. 1. This would be our predicted configuration of lowest energy. It is interesting that it is at a *non-zero* value of $r$. A value of $r=0$ would correspond to no penetration and a reduction of the La concentration in the surface plane to $f_0 = (1-x)/2$ to neutralize the surface, a very simple, though now seen to be higher energy, resolution of the surface-charge problem.

Concerning our estimate of an $r$ considerably less than the 0.74 which fit the data in Fig. 1, we may ask if at finite temperature entropy might cause a further spread of the depletion region, a question of statistical mechanics which we may address. We may readily evaluate the number $W$ of ways ions can be arranged for a particular set of $f_n$ for each plane of $N_n$ sites, and take its logarithm, to obtain $\ln(W) = -N_p \Sigma_n[f_n \ln(f_n)+(1-f_n)\ln(1-f_n)]$. This is to be minimized with respect to the $f_n$ subject to the conditions that the total number $\Sigma_n N_p f_n$ of La ions be fixed and the energy $E_{tot} = N_p/2 \Sigma_n Z^*e\phi_n f_n$ be fixed, applied with Lagrange multipliers $\mu$ and $\beta$. This would be simple if the $Z^*e^2\phi_n$ were fixed as in ordinary statistical problems, but in this interacting-many-body problem the potentials $\phi_n$ depend directly on the many other $f_n$ so we cannot simply take a derivative and obtain the usual $f_n$ given by $f_n/(1-f_n) = \exp(-\beta(Z^*e^2\phi_n - \mu))$. It is not even close to correct. We can, however, pick a form such as our $f_n = (-1)^n(1-x)(1-r^n)$ to obtain both the $\ln(W)$ and the $E_{tot}$ and then vary the $r$ to obtain the maximum $\ln(W)$ subject to a fixed energy and number of La ions. We found that in fact increasing the temperature led to even smaller $r$, depletion closer to the surface. The depletion $(1-r^n)(1-x)$ itself led to a contribution to $\Sigma_n \ln[f_n/(1-f_n)]$ decreasing with increasing $r$, as did the transfer of La to the bulk, and back to $n=0$. It does not help the discrepancy, but it may be an interesting prediction that increased temperature *decreases* the depth from the surface of the depletion. We conclude that our underestimate of $r$ must come from other terms in the energy which we have not included.

One contribution comes immediately to mind. We may imagine that the total electrostatic energy of *bulk* LSM, or perhaps the bonding energy though we have argued that it does not vary much with La rearrangement, varied as some $E(f)$, with $f = 1-x$ equal to the fraction of sites La. If this were linear in $f$ then the alloy would have the same energy as a fraction $f$ of pure $LaMnO_3$ and a fraction $1-f$ of pure $SrMnO_3$, and rearranging the ions would not affect that energy. However, it will not be linear and this gives a contribution to the energy when we redistribute the La ions. We have already included one such contribution since for the bulk crystal, $r=0$ in Eq. (3), the energy varies as $(\pi Z^{*2}e^2/8d)f^2$, quadratic in $f$, but this is only the contribution from non-neutrality of the *planes*. We may recalculate the *full* Madelung energy for the bulk crystal, using the effective charges obtained in Appendix B, and subtract that partial contribution, to obtain the correction to the bulk energy as a function of $f$. We do this in Appendix B. The resulting Madelung energy in fact has curvature as a function of $f$ of opposite sign to the $(\pi Z^{*2}e^2/8d)f^2$ and adding the difference does increase the spread to $r = 0.41$, still short of the observed 0.72. That estimate was shown as the dashed line in Fig. 1.

Again, we have only included electrostatic contributions to the energy, which are the terms which required some form of surface neutralization in the first place. It would seem to give compelling evidence that the surface charging is the origin of the segregation of strontium near the surfaces of LSM, though with only these electrostatic contributions we have not obtained a quantitative prediction of the depth of segregation, or lanthanum depletion which seems a more apt description.

Finally we might consider this mechanism in terms of other systems. First we see that it is not expected in a system such as yttrium-stabilized zirconia (YSZ). We look first at pure zirconia in the same cubic structure, a face-centered-cubic lattice of $Zr^{4+}$ ions, with $O^{2-}$ ions at the center of each of the eight "cubies", making up the face-centered cube (the fluorite structure). (100) faces would be highly charged, made entirely of Zr ions or made entirely of O ions, and such chargede surfaces would not be expected to occur. However, (110) faces would each contain one Zr for every two O ions and would be neutral. We expect (110) surfaces to be the natural cleavage and growth surfaces for this structure. If a fraction of the Zr ions are replaced with $Y^{3+}$ ions, the system is known to maintain charge neutrality by forming one oxygen vacancy, $V_O^{2+}$, for every two yttrium substitutions. Then for a uniform distribution of yttrium ions and oxygen vacancies each (110) plane acquires one vacancy for each two yttrium ions and remains neutral. The difference from LSM is that each plane has the full crystal stoichiometry, so bulk charge neutrality leads to planar neutrality. The same would be true of ceria which also occurs in the fluorite structure. It would seem that this segregation only occurs in more complex structures such as the perovskite structure of LSM. In this structure there seem not to be planes containing the $ABO_3$ stoichiometry (at least no low-index planes, which are favored). Even in this perovskite structure, (100) planes will maintain formal neutrality as long as the formal valence of the A land B are of same, three for these *oxide* perovskites. With any substitution of different valence, charged planes arise and segregation of the same form as for LSM may be expected.

## Acknowledgement

This work was supported by the National Energy Technology Laboratory under Contract No. PPM 300.02.08 with Leonard Technology, Inc.

## Appendix A  The Electrostatic Energy

We start with a set of planes numbered as in Fig. 1 with charges

$$eZ^*f_n = eZ^*(-1)^n(1-x)(1-r^n) \ . \tag{A1}$$

The first task is to calculate the total charge (in units of $eZ^*$ per $4d^2$) $Q_n = \Sigma_{j=0, n-1} f_n$ for planes numbered $j < n$. After summing a geometric series, we obtain exactly

$$Q_n = (1-x)(r-r^n)/(1+r) \quad \text{for odd } n,$$
and (A2)
$$Q_n = -(1-x)(1-r^n)/(1+r) \quad \text{for even } n.$$

We may note immediately that this does not correspond at large $n$ to a $\pm Q^*$ as it would if we had completely eliminated the surface charge by shifting La from the surface. A surface charge remains equal to $-\frac{1}{2}(1-x)(1-r)/(1+r)$. We could eliminate it by changing that number of $Mn^{3+}$ ions to $Mn^{4+}$ at the plane $n = 1$. Then if $r$ were near zero, eliminating the spread of depletion, this becomes exactly the promotion of $\frac{1}{2}(1-x)$ which we assumed in Ref. 6 for $MnO_2$ surfaces. An alternative, which we find has lower energy, is to add an La ion at a fraction $\frac{1}{2}(1-x)(1-r)/(1+r)$ of the A sites in the $n=0$ plane. This then leads to

$$Q_n = (-1)^n (1-x)[-1/2 + r^n/(1+r)] \quad \text{for all } n. \tag{A3}$$

Such a planar charge, with no field in the space to the left of $n = 0$ in Fig. 1 yields a $e\partial\phi/\partial x$ just to the left of the $j$th plane of $-(\pi e^2/d^2)Q_j$. Multiplying that by the spacing $d$ gives the shift in energy between the two planes, so when we add that shift to all the shifts for planes with smaller $j$ than $n$ we obtain $e\phi_n = -(\pi Z^* e^2/d)\Sigma_{j=0,n-1} Q_j$. After another sum of a geometric series we obtain

$$e\phi_n = (-1)^n (\pi Z^* e^2/d)(1-x)[1/4 - r^{n+1}/(1+r)^2]. \tag{A4}$$

There was an additional constant term, $-(\pi Z^* e^2/d)/(1+r)^2$, which was dropped so that $\phi_n$ is measured from the average value deep in the crystal.

From this we may calculate the electrostatic energy per surface area $4d^2$ given by

$$\tfrac{1}{2} \Sigma_n eZ^* f_n \phi_n = (\pi Z^{*2} e^2/2d)(1-x)^2 \Sigma_n (1-r^n)[1/4 - r^{n+1}/(1+r)^2]. \tag{A5}$$

The term $\frac{1}{4}$ gives $(\pi Z^{*2} e^2/8d)N_0$, the energy of a bulk crystal with no surface terms, and we drop it. The remaining terms under the sum become $-r^n/4 - r^{n+1}(1-r^n)/(1+r)^2$. Our choice of $f_n$ has reduced the total number of La ions by $\Sigma_{even\ n} (f_n - (1-x)) = -(1-x)\Sigma_{even\ n} r^n$ (minus the $\frac{1}{2}(1-x)(1-r)/(1+r)$ ions we placed in the $n=0$ plane, to which we return). [It can be seen to be equal to $-(1-x)/(1-r^2)$ per $4d^2$ if we sum over even $n$. Similarly the number of $Mn^{3+}$ is found to be reduced by $-(1-x)r/(1-r^2)$, summing over odd $n$ for a total $-(1-x)/(1-r)$. However, we wish to write all $n$-dependent terms in the energy before summing.] These La ions must be placed in the bulk where they have a positive electrostatic energy of $(\pi e^2/d)(1-x)/4$ from Eq. (A3), and the corresponding number of $Mn^{4+}$ ions will be converted to $Mn^{3+}$ ions for a total electrostatic energy contribution of $(\pi e^2/d)(1-x)^2 \Sigma_{all\ n} r^n/4$. Note that this is not divided by two, as in Eq. (A5), because it is a change in charge times the original potential; there is also a change in potential times the original charge.

This contribution is to be added to Eq. (A5) and is seen to just reverse the first of the terms which appeared under the sum, $-r^n/4 - r^{n+1}(1-r^n)/(1+r^2)$, leaving

$$½ \Sigma_n eZ^*f_n\phi_n = (\pi Z^{*2}e^2/2d)(1-x)^2 \Sigma_n [r^n/4 - r^{n+1}(1-r^n)/(1+r)^2]$$
$$= (\pi Z^{*2}e^2/2d)(1-x)^2 [1/(4(1-r)) - r^2/((1-r)(1+r)^3)], \tag{A6}$$

completing the calculation of the electrostatic energy, with the first form associating the total with individual planes. Finally we include the energy associated with placing $½(1-x)(1-r)/(1+r)$ La in the $n=0$ plane, where their potential is lower than the bulk value by $-(\pi Z^{*2}e^2(1-x)/d)r/(1+r)^2$ from Eq. (A4). The total change in energy associated with the redistribution of La ions is becomes

$$E_{tot.} = (\pi Z^{*2}e^2/2d)(1-x)^2\{-r^2/[(1-r)(1+r)^3] + 1/[4(1-r)] - r(1-r)/(1+r)^3\}. \tag{A7}$$

This is plotted in Fig. A1. We see that the minimum energy occurs at quite a small value of $r = 0.19$ where the energy was $0.191(\pi Z^{*2}e^2/2d)(1-x)^2 = 0.37$ eV. Both there and at $r = 0$, corresponding to no spread in the depletion and energy 0.49 eV, it is well below the ½ $(1-x)E_{pro} = 0.83$ eV which would be required by promoting Mn ions to neutralize the surface.

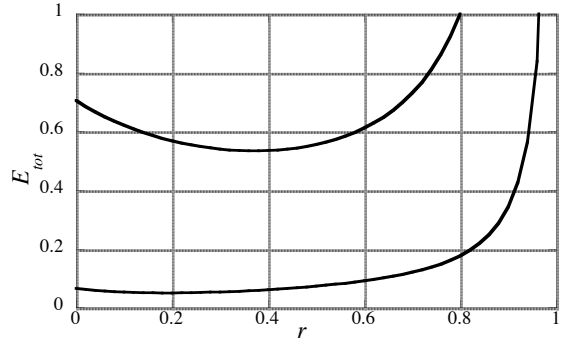

Fig. A1. The total energy in units of $e^2/d$, per area $4d^2$ as a function of $r = e^{-\mu dn}$. The lower curve is from Eq. (A7), the upper has added Madelung terms from Eq. (B4).

## Appendix B  Effective Charges and Madelung Energies

We have associated an effective charge $Z^*$ with the bulk LSM planes. We calculated such charges for a wide range of solids in Ref. 8. There was no direct way to test these values but the measurable transverse charges $e_T^*$ and the dielectric susceptibility which were obtained as an extension of the $Z^*$ calculations agreed quite well with experiment. These calculations did not include the manganites so we did them now using exactly the same approach and parameters from Ref. 7. The calculations start with +2 for $Sr^{2+}$, for example, and form cluster orbitals for the Sr $s$, $p$, and $d$ states, all of which are empty in the free 2+ ion (but coupled to neighboring oxygen $p$ states). The share of each of these orbitals on the Sr is calculated and subtracted from the +2 for this electronic charge placed on the ion.

The calculations assumed the bulk spacing for $LaMnO_3$ and $SrMnO_3$ which may be appropriate since in the alloy the neighbors relax from their averages spacing toward the pure-material spacing (Ref. 5, 296ff). For $LaMnO_3$ we obtained $Z^*(La) = 1.92$, $Z^*(Mn^{3+}) = 1.16$, and $Z^*(O) = -1.03$. For $SrMnO_3$ we obtained $Z^*(Sr) = 1.29$, $Z^*(Mn^{4+}) = 1.45$, and $Z^*(O) = -0.91$. Then for $La_{0.65}Sr_{0.35}MnO_3$ we use the $Z^*(La)$ and $Z^*(Sr)$ and the weighted average of $Z^*(O) = -0.99$. Then $Z^*$ is the difference when and La is substituted for an Sr, the charge we seek, $Z^* = Z^*(La) - Z^*(Sr) = 0.63$.

We turn next to the full Madelung calculation. It was made for general charges in the cubic perovskite structure in Ref. 9. We may combine terms from that paper, noting that $Z^*(O) = -(Z^*(A)+Z^*(B))/3$, to obtain the total electrostatic energy per formula unit of

$$E_{tot} = \tfrac{1}{2}\Sigma Z_j^* e\phi_j = -1/2\,[1.51Z^*(A)^2 + 1.46Z^*(A)Z^*(B) + 1.99Z^*(B)^2]e^2/d. \qquad (B1)$$

We may check that it is close to the accepted[9] $-24.76 e^2/d$ for $SrTiO_4$ with formal valences of 2, 4, and $-2$. For this analysis we write $g_n$ as the fraction of Sr or $Mn^{4+}$ sites, going to $x$ in the bulk. [The fact that $f_n$ fluctuates in sign and goes to $1-x$ in the bulk proved confusing in this context.] Then also a fraction $g_n$ of the B sites are $Mn^{4+}$ and we have $Z^*(A) = (1-g_n)Z^*(La) + g_n Z^*(Sr)$ and $Z^*(B)=(1-g_n) Z^*(Mn^{3+}) + g_n Z^*(Mn^{4+})$. Using the effective charges given above, $E_{tot}$ varies from $-3.69e^2/d$ at $g_n = 0$ to $-6.91e^2/d$ at $g_n = 1$. More importantly, if we subtract the linear interpolation between the two limits we obtain

$$E_{tot}(g_n) = 0.517(0.673)\, g_n (1-g_n)e^2/d, \qquad (B2)$$

with the (0.673) the corrected value we shall find. If we take the corresponding bulk energy from Eq. (A7) by setting $r = 0$, and subtract the linear interpolation we obtain $E_{tot} = -(\pi Z^{*2}e^2/8d)\, g_n(1-g_n) = -0.156e^2\, g_n(1-g_n)/d$, which we subtract from Eq. (B2) as already included in our calculation, though that calculation also included the electrostatic potential from distant charged planes of different $g_n$ which are retained. Then the leading coefficient in Eq. (B2) is increased to 0.673 as indicated in parentheses.

We now let the $g_n$ represent the local Sr or $Mn^{4+}$ concentration, always positive, and varying with position as $x+(1-x)r^n$. We may readily substitute this in Eq. (B2) (with the new coefficient) and sum over $n$ to obtain an $r$-dependent contribution to the total energy

$$E_{tot}(\text{Madelung}) = 0.673(e^2/d)[x(1-x)N_0 + (1-x)(1-2x)/(1-r) - (1-x)^2/(1-r^2)]. \qquad (B3)$$

The first term is the bulk energy without surface effects, and not of interest. Again, this has eliminated $(1-x)/(1-r)$ ions (La and $Mn^{3+}$) to be placed in the bulk at $\partial E_{tot}(g_n)/\partial g_n|_x = 0.673(e^2/d)(1-2\,g_n)|_x$, just canceling the second term in Eq. (B3). Finally we place $g_0 = \tfrac{1}{2}(1-x)(1-r)/(1+r)$ La ions in the $n=0$ plane from the bulk and note that Eq. (B2) is the same for $f_0$ replacing $g_0=1-f_0$ but $\partial E_{tot}(f_n)/\partial f_n|_{1-x} = -\partial E_{tot}(g_n)/\partial g_n|_x$.

Then $E_{tot}(f_0) - f_0 \partial E_{tot}/\partial f|_{1-x} = 0.673(e^2/d) f_0 [2-2x-f_0]$, for a total $r$-dependent energy from the variation with $x$ in the full bulk Madelung energy of the LSM of

$$E_{tot}(\text{Madelung}) = 0.673(e^2/d)[(1-x)^2/(1-r^2) + f_0(2(1-x)+f_0)], \tag{B4}$$

with $f_0 = \frac{1}{2}(1-x)(1-r)/(1+r)$.